**White Paper on Opportunities in Plasma Physics**

Submitted to The National Academy of Sciences, Engineering, and Medicine
in response to the 2020 Decadal Study on Plasma Physics

# Laser-Plasma Interactions Enabled by Emerging Technologies

Author: John Palastro,
Institution: University of Rochester, Laboratory for Laser Energetics
Email: jpal@lle.rochester.edu
Phone: (585) 275-9939

Co-Authors:
Felicie Albert[1], Brian Albright[2], Thomas Antonsen Jr.[3], Alexey Arefiev[4], Jason Bates[5], Richard Berger[1], Jake Bromage[6], Michael Campbell[6], Thomas Chapman[1], Enam Chowdhury[7], Arnaud Colaïtis[8], Christophe Dorrer[6], Eric Esarey[9], Frederico Fiúza[10], Nathaniel Fisch[11], Russell Follett[6], Dustin Froula[6], Siegfried Glenzer[10], Daniel Gordon[5], Daniel Haberberger[6], Bjorn Manuel Hegelich[12,13], Ted Jones[5], Dmitri Kaganovich[5], Karl Krushelnick[14], Pierre Michel[1], Howard Milchberg[3], Jerome Moloney[15], Warren Mori[16], Jason Myatt[17], Philip Nilson[6], Steve Obenschain[5], Jonathan Peebles[6], Joe Peñano[5], Martin Richardson[18], Hans Rinderknecht[6], Jorge Rocca[19], Andrew Schmitt[5], Carl Schroeder[9], Jessica Shaw[6], Luis Silva[20], David Strozzi[1], Szymon Suckewer[11], Alexander Thomas[14], Frank Tsung[16], David Turnbull[6], Donald Umstadter[21], Jorge Vieira[20], James Weaver[5], Mingsheng Wei[6], Scott Wilks[1], Louise Willingale[14], Lin Yin[2], Jon Zuegel[6]

Institutions:
[1]Lawrence Livermore National Laboratory
[2]Los Alamos National Laboratory
[3]University of Maryland, College Park
[4]University of California, San Diego
[5]Naval Research Laboratory
[6]University of Rochester, Laboratory for Laser Energetics
[7]Ohio State University
[8]Université de Bordeaux, Centre Lasers Intenses et Applications, France
[9]Lawrence Berkeley National Laboratory
[10]SLAC National Accelerator Laboratory
[11]Princeton University
[12]Gwangju Institute of Science and Technology, Center for Relativistic Laser Science,
[13]University of Texas, Austin
[14]University of Michigan
[15]University of Arizona
[16]University of California, Los Angeles
[17]University of Alberta
[18]University of Central Florida
[19]Colorado State University
[20]Instituto Superior Técnico, Portugal
[21]University of Nebraska, Lincoln

**Introduction:** The recent awarding of the Nobel Prize to Donna Strickland and Gerard Mourou for chirped pulse amplification (CPA) has highlighted the impact that broadband laser systems have had throughout science [1]. Plasma physics, in particular, has developed a unique synergy with CPAs: plasma has provided the only medium that can withstand the increases in intensity delivered by CPAs over the last 30 years. CPA is not, however, the only breakthrough in optics technology that has or promises to expand the frontier of laser-plasma interactions. Over the past ten years, creative optical solutions have produced unprecedented intensities, contrast, repetition-rates, and gain bandwidths, renewed interest in long wavelength drivers, and provided novel methods for spatiotemporal pulse shaping. Over the next 10 years, these emerging technologies will advance diverse fields of plasma physics, including:

- Direct drive, indirect drive, and magnetized inertial confinement fusion
- Nonperturbative nonlinear propagation and material interactions
- Advanced accelerators
- Plasma-based radiation sources
- Ultrahigh magnetic field generation
- High-field and electron-positron plasmas
- Structured light-plasma interactions

One such field—*plasma optics*—promises to overcome fundamental limitations of solid-state optical technology and will usher in the next frontier of plasma research. Progress and science discovery in pursuit of this frontier will require a community approach to experiments, simulations, and theory, together with investments in an ecosystem of laser facilities and simulation software.

**Inertial Confinement Fusion:** With the global population rising to over 9 billion by the end of the 21$^{st}$ century and the rising tide of climate change, the pursuit of environmentally acceptable energy sources has become more critical than ever. While still in the research stage, controlled fusion could deliver an almost endless supply of power with relatively low environmental impact. The inertial confinement fusion (ICF) approach, in particular, was one of the earliest applications to harness high-power lasers and has, on several occasions, implemented innovative optical techniques to effect step-changes in performance: efficient frequency tripling, spatial coherence control (phase-plates), and induced temporal incoherence (ISI and SSD) [2-6]. While these successes have allowed ICF to push the intensity ever higher, designs must still navigate around laser-plasma instabilities and laser-imprint [7,8]. Laser-plasma instabilities inhibit the deposition of energy in the ablator and put the laser at risk for damage by scattering light into unwanted directions. Moreover, these instabilities can generate super-thermal electrons that preheat the fusion fuel, reducing its compressibility. Laser-imprint, i.e. density non-uniformities on the capsule surface imparted by speckles, seeds the Rayleigh-Taylor instability and can cause the capsule to break up during compression.

Creative uses of the bandwidth available on current laser systems may inhibit low-frequency laser-plasma instabilities like stimulated Brillouin scattering by detuning the interaction between multiple laser beams or by moving speckles before the instability can grow [9]. **Modern broad-bandwidth lasers, on the other hand, could revolutionize ICF by providing** **unprecedented spatiotemporal control over laser-plasma interactions.** These lasers can deliver pulses with the temporal incoherence necessary to suppress high frequency instabilities like two-plasmon decay and stimulated Raman scattering, while also providing smoothing sufficient to eliminate imprint [10-12]. Generally speaking, the broad-bandwidth mitigates laser plasma instabilities by detuning the interaction between multiple waves or incoherently drives many small instabilities instead of a single coherent instability. To this end, optical parametric amplifiers (OPAs) offer an excellent candidate for the next generation ICF driver. OPAs create high power, broad bandwidth light that can be seeded with a variety of temporal formats, including the random intensity fluctuations of parametric fluorescence, spike trains of uneven duration and delay (STUDs), or chirplets [13]. The bandwidth of the resulting pulses, or those from an existing wideband architecture, could be further increased by stimulated rotational Raman scattering (SRRS) during propagation to the final focusing optics. Preliminary experiments have demonstrated this technique can broaden the spectrum of frequency multiplied Nd:glass and KrF pulses to multi-terahertz bandwidths [14].

**Nonperturbative nonlinear propagation and material interactions:** The field of nonperturbative nonlinear propagation and material interactions spans the boundary of nonlinear optics and plasma physics with relevant dynamics occurring over a trillion orders of magnitude in time: starting with the attosecond dynamics of bound electrons that determine the nonlinear optical response, evolving into the femto and picosecond formation and evolution of plasma, and concluding with the micro and millisecond hydrodynamic evolution of the neutral medium. **The emergence of high-power, high-repetition-rate (>kHz) ultrashort pulse lasers enables novel regimes of nonlinear propagation and material interactions governed by a combination of non-thermal and thermal modifications to matter**—regimes with scientific, industrial, and security applications such as understanding new states of warm dense matter, femtosecond micromachining, laser eye surgery, electromagnetic pulse (EMP) generation, and long-range propagation through atmosphere for remote sensing.

For a single high-power pulse propagating in transparent media, plasma formation counteracts the nonlinear collapse during self-focusing, leading to high-intensity propagation over distances much longer than a Rayleigh range [15-18]. At high rep-rate, each laser pulse experiences a nonlinear environment modified by its predecessors, which combines traditional effects such as thermal blooming [19] with ultrafast nonthermal effects, including ionization and impulsive (molecular) Raman excitation [20]. Already experiments have demonstrated that a train of laser pulses can heat air through these processes, leaving behind a long-lasting neutral density channel that can guide subsequent laser pulses [21] and enhance the collection efficiency in remote detection [22].

For high-rep rate material interactions, a laser pulse will interact with matter that has been strongly modified by the non-thermal heating of previous pulses. This heating can create periodic surface structures, change the reflectivity and absorption, or alter the molecular composition altogether. The interaction involves multiple physics phenomena, including the time-dependent dielectric response, stimulated scattering mechanisms, phase changes, electronic bandgap structure, and combined optical-collisional photoionization. In many of the solid and liquid media relevant to applications, the material properties governing these phenomena are not well characterized or even measured. Expanded use of spectral interferometry measurements [23,24], as well as the pursuit of new techniques, could greatly improve understanding and facilitate the development of applications.

**Plasma Accelerators:** Particle accelerators provide a looking glass into a sub-atomic world inhabited by the fundamental building blocks of the universe. Conventional accelerators, based on vacuum technology, continue to make impressive strides, routinely improving beam quality and achieving unprecedented energies. With each advance, however, conventional accelerators grow in size or cost. Laser-plasma accelerators promise to break this trend by taking advantage of the extremely large fields either inherent to or driven by ultrashort laser pulses and a medium—plasma—that can sustain them. Armed with a vision of smaller-scale, cheaper accelerators and empowered by advances in laser technology, these "advanced accelerators" have achieved rapid breakthroughs in both electron and ion acceleration.

Early laser-wakefield acceleration (LWFA) experiments made steady progress by trapping and accelerating electrons in plasma waves excited by unmatched laser pulses—pulses with durations exceeding the plasma period [25,26]. Such pulses confined LWFA to sub-optimal regimes in which plasma waves were driven either by laser pulse self-modulation or beat waves. With the advent of high-power, broadband multi-pass amplifiers, progress exploded—to this day, the maximum electron energy continues to climb with laser power [27-30]. These amplifiers deliver ultrashort pulses with durations less than the plasma period, allowing experiments to access the forced, quasi-linear, and bubble regimes [31-33]. Aside from increasing the maximum energy of the electron beams, the ultrashort pulses enable transformative injection techniques, through self-trapping or controlled ionization, that greatly reduced the electron beam emittance and energy spread [34-36]. **The emergence of amplifiers that can operate at both high peak and high average power provide a technological path towards a LWFA-based electron-positron collider.** While many physics and technology challenges must still be overcome, the high-rep rates of these systems could deliver the luminosity needed to achieve a number of events comparable to traditional colliders [37].

While the large inertia of ions precludes their efficient acceleration through LWFA, a high-intensity pulse incident on a solid or shocked target can drive several mechanisms that accelerate ultrashort, high-flux ion beams from rest [38-45]. These mechanisms can be broadly separated into a few categories: ions accelerated by the sheath field of hot electrons escaping the back side of a solid target, or target-normal sheath acceleration [38,39]; ions gaining energy by reflecting off a moving electrostatic potential, either caused by radiation pressure (hole-boring) [40] or thermal pressure (shock acceleration) [41,42]; beam-plasma modes excited during relativistic transparency [43]; solitary wave generation [44]; or hybrid schemes that combine elements of these with other mechanisms. Developments in high-power, broadband amplifiers have made sources based on these mechanisms widely accessible for a range of applications, producing ion beams with energies comparable to longer, higher energy pulses. Proton sources, for instance, are now routinely used to radiograph high-energy-density matter, providing an invaluable probe for resolving plasma dynamics on picosecond time scales [38,46]. **Advances in laser contrast and amplifiers that operate at both high peak and high average power would represent a transformative step towards the realization of laser-driven proton/ion beams as injectors for high-brilliance accelerators and medical therapy.** When integrated with recent developments in high-rep rate cryogenic targets, high-rep rate lasers offer significantly greater control over the acceleration process and enable high-quality beams with 10s to 100s MeV and high particle flux. This integration would also provide an ideal platform for understanding the origin and evolution of magnetic instabilities in proton beams [45].

**Radiation sources:** The strong accelerations experienced by electrons in intense laser-plasma interactions unleashes a torrent of secondary radiation that spans the electromagnetic spectrum. **Leveraging increases in laser rep-rate and intensity with creative interaction configurations and plasma structuring could spark the development of plasma-based radiation sources that excel in throughput, brightness, coherence, power, or efficiency.** These plasma-based sources offer compact, low-cost alternatives to sources based on conventional accelerators that, if harnessed, could be widely accessible for applications.

The development of sources in two frequency bands in particular, x-ray and terahertz (THz), would have far-reaching benefits in medicine, defense, and basic science [47]. Laser-plasma interactions generate x-rays through a number of diverse mechanisms: betatron radiation from electrons oscillating in wakefields [48,49], bremsstrahlung emission from energetic electrons crashing into high-density matter [50], laser photons double Doppler-upshifted by a counter-traveling, relativistic electron beam, i.e. Compton scattering [51,52], stimulated emission of photons from relativistic electrons wiggling in a free electron laser [53], x-ray lasing through transient collisional excitation [54,55], or high harmonic generation from electrons accelerated in and out of a surface by an intense laser-field [56]. This diversity provides the flexibility to choose a mechanism that best meets the requirements of applications such as phase contrast imaging, radiosurgery, lithography, and nuclear resonance fluorescence for standoff detection of radioactive or other threatening materials. On the opposite end of the spectrum, the interaction of intense laser pulses with structured plasmas can efficiently drive THz radiation. The ponderomotive force of a laser pulse excites a time-dependent current. In a non-uniform plasma, this current radiates into the far-field, emitting frequencies within a band determined by the pulse duration [57,58]. This radiation could bridge the "terahertz gap"—the scarcity of sources between the frequency ranges accessible by electronics and lasers—and do so with high power, ultrashort THz pulses. In contrast to x-rays, THz radiation is non-ionizing and can be safely used for non-invasive biomedical imaging and medical tomography. Further, the energy separation of rotational-vibrational eigenstates makes THz radiation ideal for time domain spectroscopy and standoff detection of chemical and biological molecules [59,60]. In terms of discovery science, THz radiation can directly excite matter to highly excited phonon states, unlocking new regimes of high-energy density physics [61].

**Magnetized plasmas:** Like plasmas, magnetic fields occur ubiquitously throughout the Universe and play a critical role in shaping astrophysical environments. Emulating these environments in the laboratory with well-diagnosed experiments can provide a valuable complement to conventional astrophysical observations. High-power lasers facilitate these experiments by creating scale-equivalent plasma conditions with self-generated or external magnetic fields, or by directly driving up magnetic fields through laser-plasma interactions. Either way, the magnetic fields fundamentally alter the laser-plasma interaction. The presence of ultra-strong, quasi-static magnetic fields modifies the microscopic kinetics by diverting, confining, or undulating electrons, the collective behavior by bringing the cyclotron resonance within reach of optical excitation, and laser propagation through peculiar dispersive effects such as polarization rotation, slow light, and induced transparency.

The capability to perform controlled, focused experiments by generating strong magnetic fields with lasers has only recently emerged. The current approach, based on existing laser technology, uses a long, high energy pulse to drive a current through induction coils. Aside from basic laboratory astrophysics, these platforms allow for investigations into magnetized high-energy density physics related to the transport of high energy particles and high-gain ICF schemes like fast-ignition[62]. **The projected intensities delivered by next-generation laser systems could directly drive volumetric magnetic fields rivaling those occurring on the surface of neutron stars (~MT).** These extreme fields, created by the highly-nonlinear currents driven by an intense laser pulse propagating through a relativistically transparent, high-density plasma, would result in a number of immediate breakthroughs [63-65]: they would significantly enhance the transfer of energy from a laser pulse to electrons and facilitate the emission of gamma rays from relativistic electrons by providing a powerful undulator [64,66]. The development of such a gamma ray source would be critical for the development of nuclear and radiological detection systems. Furthermore, the gamma-ray source would enable discoveries linked to our understanding of the early Universe and high-energy astrophysics, including the direct creation of matter and anti-matter from light [67] and allow the direct control and study of nuclear excitation and structure [68].

**High Field and Electron-Positron Plasmas:** Nonperturbative QED represents the current frontier of laser-plasma interactions—a frontier in which ripping electron-positron pairs from the Dirac sea may make targets a thing of the past [69]—a frontier in which vacuum exhibits magnetization, polarization and birefringence [67]—a frontier in which the analogy of Hawking radiation in electric fields, Unruh radiation, could provide insight into the life-cycle of black holes [70]. Compared to any other physical theory, perturbative QED predictions have been experimentally confirmed to unprecedented accuracy. While electric fields strong enough to accelerate an electron to its rest mass over a Compton wavelength, i.e. at the Schwinger limit, are sufficient to test nonperturbative QED models, creative laser-plasma configurations can create highly nonlinear environments at much lower field strengths. This strategy has already proven successful in experimental demonstrations of nonlinear Compton scattering [71], positron production [72,73], and radiation reaction [74]. Nevertheless, the exotic theoretical and computational predictions of nonperturbative QED models have rapidly outpaced the experimental capabilities to test them. **By providing flexible laser-plasma configurations and extremely high intensities, a next-generation laser could access unexplored regimes of nonperturbative, collective QED effects in plasmas and test the exotic predictions of the models.** Such a facility could bring the mysteries of astrophysical objects, including black holes, pulsars, and magnetars, down to earth, and uncover the dynamic interaction of inner shell electrons with highly ionized, heavy nuclei [75,76].

**Structured light-plasma interactions:** Beyond simply adjusting parameters like intensity and frequency, the spatiotemporal structure of light offers additional degrees of freedom for controlling the interaction of intense laser pulses with plasma. Structured light fields emerge spontaneously when two or more electromagnetic plane waves interfere. The interference of three waves, for instance, can produce phase singularities, which give rise to one of the most fascinating features of structured light: orbital angular momentum (OAM). OAM pulses can impart angular momentum to the plasma, modifying the topology and dispersion of driven waves and the phase space of the charged particles they accelerate [77,78]. As an example, a laser pulse with a helical intensity profile, or "light spring," can ponderomotively excite a wakefield that traps and accelerates a vortex electron beam, i.e. a beam that rotates around the optical axis [78]. OAM can also modify the nonlinear propagation and interaction of high-power pulses with transparent media, resulting in helical plasma filaments or high harmonic radiation with vortex phase structure [79,80].

More complex interference patterns exhibit striking properties that appear to violate special relativity: the peak intensity of a self-accelerating light beam follows a curved trajectory in space [81], while the peak intensity of a "flying focus" pulse can travel at an arbitrary velocity, surpassing even the vacuum speed of light [82]. These arbitrary velocity intensity peaks result from the chromatic focusing of a chirped laser pulse. The chromatic aberration and chirp determine the location and time at which each frequency component within the pulse comes to focus, i.e. reaches its peak intensity, respectively. By adjusting the chirp, the velocity of the intensity peak can be tuned to any value, either co- or counter-propagating along the laser axis. This, in turn, grants control over the velocity of an ionization front or ponderomotive force—a control with the potential to advance several plasma-based applications, including Raman amplification, photon acceleration, wakefield acceleration, and THz generation [83,84]. While these unexpected features of structured light bring about new and rich laser-plasma interactions, they have remained relatively unexplored due to the technological challenges of creating such pulses. **The further development of ultrafast pulse shaping techniques to manipulate the spatiotemporal degrees of freedom would provide a virtual forge for creating pulses to optimize or bring about novel laser-plasma interactions.** In doing so, these techniques would enrich all of the subfields discussed above.

**Plasma Optics:** Ultimately, advances in plasma physics will require rep-rates or intensities that exceed the damage limitations of solid-state optical components. Even with improvements in high-damage optical coatings, the size of solid-state optical components must increase to maintain tolerable fluences. Aside from the prohibitive cost of such large optics, this approach will eventually become counteractive: larger optics can withstand higher powers, but their fabrication introduces surface aberrations that reduce focusability and, as a result, the peak intensity. **Plasma-based optical components could provide the disruptive technology needed to usher in the next frontier of plasma research**. Plasma optics, being already ionized, have substantially higher damage thresholds than solid state components and can be inexpensively and rapidly replaced, for instance, at the rep-rate of a gas jet or capillary, or flow rate of a water jet [85-89].

Similar to conventional optics, a laser pulse propagating in plasma acquires a spatiotemporal phase determined by the refractive index. By controlling the spatial variation, evolution, or nonlinearity of the plasma density, the plasma can provide dispersion, refraction, or frequency conversion, respectively, and, in principle, be made to mimic any solid-state optical component. Already, several such components routinely improve experimental performance: plasma gratings successfully tune the implosion symmetry of ICF capsules at the National Ignition

Facility [90]; plasma waveguides combat diffraction, extending the interaction length in LWFAs [91]; and plasma mirrors (1) enhance intensity contrast by orders of magnitude, allowing for impulsive laser-matter interactions free of premature heating [92] and (2) redirect laser pulses in multi-stage LWFAs without degrading electron beam emittance [93]. Several other plasma components, while still in the nascent stages of development, have been successfully demonstrated in experiments: lenses [94,95], waveplates [96], q-plates [97], beam-combiners [98], compressors, and amplifiers [99]. Plasma amplifiers, in particular, could eventually replace CPAs in the final power-amplification stage of a laser, eliminating the need for large, expensive gratings [100]. In principle, these amplifiers can achieve intensities $10^3$ times larger than CPAs in the infrared or operate in wavelength regimes inaccessible to CPAs altogether, e.g. the ultraviolet or x-ray range [101]. A next-generation high-power laser that implemented plasma components could deliver extremely high intensity pulses—pulses that would transform the landscape of laser-plasma interactions.